\documentclass[twocolumn,aps,showpacs,preprintnumbers,superscriptaddress]{revtex4}
\usepackage{graphicx}
\usepackage{dcolumn}
\usepackage{bm}

\begin{document}

\title{Squeezing and entanglement of matter-wave gap solitons}

\author{Ray-Kuang Lee}
\affiliation{Nonlinear Physics Centre and ARC Centre of Excellence for Quantum-Atom Optics, Research School of Physical Sciences and Engineering, The Australian National University, Canberra, ACT 0200, Australia}
\affiliation{Department of Photonics and Institute of Electro-Optical Engineering, National Chiao-Tung University, Hsinchu 300, Taiwan}

\author{Elena A. Ostrovskaya}
\affiliation{Nonlinear Physics Centre and ARC Centre of Excellence for Quantum-Atom Optics, Research School of Physical Sciences and Engineering, The Australian National University, Canberra, ACT 0200, Australia}

\author{Yuri S. Kivshar}
\affiliation{Nonlinear Physics Centre and ARC Centre of Excellence for Quantum-Atom Optics, Research School of Physical Sciences and Engineering, The Australian National University, Canberra, ACT 0200, Australia}

\author{Yinchieh Lai}
\affiliation{Department of Photonics and Institute of Electro-Optical Engineering, National Chiao-Tung University, Hsinchu 300, Taiwan}

\begin{abstract}
We study quantum squeezing and entanglement of gap solitons in a Bose-Einstein condensate loaded into a one-dimensional optical lattice. By employing a linearized quantum theory we find that quantum noise squeezing of gap solitons, produced during their evolution, is enhanced compared with the atomic solitons in a lattice-free case due to intra-soliton structure of quantum correlations induced by the Bragg scattering in the periodic potential. We also show that nonlinear interaction of gap solitons in dynamically stable bound states can produce strong soliton entanglement.  
\end{abstract}

\maketitle
%Introduction (new)
In the past two decades, the active research into quantum properties of nonlinear many-particle systems has enabled control and engineering of quantum noise in nonlinear optics \cite{review}. Quantum noise reduction of optical signals below the shot-noise level, referred to as uncertainty squeezing, and quantum entanglement of interacting optical pulses are at the core of the applications of non-classical light in quantum interferometry, precision measurement, and quantum information processing. Non-spreading optical collective excitations, {\em solitons}, supported by dispersive nonlinear media, proved the best candidates for experiments on quantum noise reduction and entanglement due to their robust dynamics and scattering properties \cite{review}.

It has been recognized \cite{werner_prl} that general methods of quantum noise squeezing and entanglement developed in quantum optics could apply to other nonlinear  bosonic fields, such as weakly interacting ultracold atoms in a Bose-Einstein condensate (BEC). Consequently, a number of theoretical proposals were put forward \cite{Pu00,Duan,Kuang} and experiments carried out \cite{Orzel01,Sorensen01,Ketterle} on generating macroscopic entangled number-squeezed states in BEC. Production of quantum correlations in the macroscopic quantum states relies on interactions between the atoms, the mechanism that is analogous to optical Kerr effect.  

Recently, {\em gap solitons} in a {\em repulsive} condensate, supported by periodic potentials of optical lattices, have attracted a great deal of attention due to their controllable interaction and robust evolution uninhibited by collapse \cite{Zob_et99,Kon_Sal02,Pearl03,Eiermann04,lewenstein}.
The existence of gap solitons is a unique property of {\em nonlinear periodic systems}.
Atomic gap solitons form due to combination of an inherent nonlinearity of BEC and the bandgap structure of the matter-wave spectrum that can modify and ultimately reverse dispersion properties of the atomic wavepackets~\cite{dm}.
The current techniques for gap solitons generation suffer greatly from ``technical'' noise \cite{Eiermann04}.
Provided this problem can eventually be overcome, gap solitons may represent an attractive high-density source for atomic interferometry, quantum measurements and quantum information processing with ultracold atoms. Understanding and control of quantum noise associated with atomic gap solitons is therefore a fundamental issue which so far has not been explored. However, the studies of quantum polariton solitons in a frequency dispersive medium~\cite{rupasov} and the amplitude-squeezed optical solitons in shallow periodic media described by the couple-mode model~\cite{RK-fbg} suggest that both strong dispersion and bandgap spectrum of a nonlinear soliton-bearing system with periodically varying parameters can dramatically modify the properties of the soliton squeezing.

In this Letter, we study quantum fluctuations of matter-wave gap solitons in an optical lattice and investigate the band-gap effect on the quantum noise squeezing. We employ the soliton perturbation approach \cite{Lai89} to analyze quantum fluctuations around the soliton solutions of the Gross-Pitaevskii (GP) equation with a periodic potential.
Using this approach, we demonstrate enhancement of quantum noise squeezing effect induced by the gap soliton evolution.
Furthermore, we show that the existence of dynamically stable bound states of gap solitons provides a favorable environment for generating entangled soliton pairs.

\begin{figure}
\includegraphics[width=3.0in]{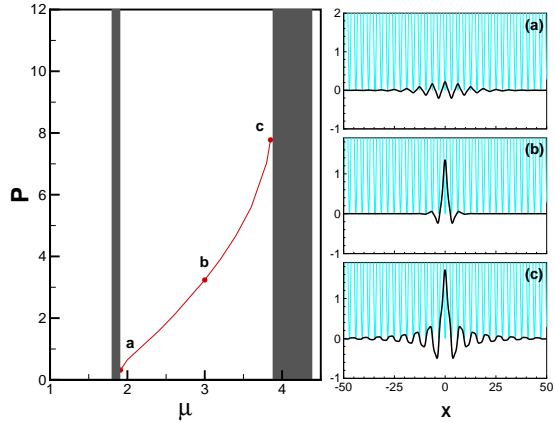}
\caption{(Color online) Left: Band-gap diagram for Bloch waves (bands are shaded), and the family of the gap solitons in the first finite gap for $V_0 = 4.0$. Right: Profiles of gap solitons with different chemical potentials; $\mu = 1.91$, $3.0$, and $3.85$, corresponding to the points $a$, $b$, and $c$, respectively.}
\label{fig_g1}
\end{figure}
We consider an elongated cigar-shape BEC loaded into a one-dimensional optical lattice and described by the GP equation for the macroscopic wave function \cite{Pearl03},
\begin{eqnarray}
i\hbar\frac{\partial \Psi}{\partial t} = -\frac{1}{2}\frac{\partial^2 \Psi}{\partial x^2}+V(x)\Psi+g_{1D}|\Psi|^2\Psi,
\label{eq_GP}
\end{eqnarray}
where $\Psi$ is the normalized macroscopic wave function of the condensate, $g_{1D}$ is the nonlinear interaction coefficient, and $V(x) = V_0\sin^2(x)$ is the one-dimensional periodic potential. Stationary states of the condensate can be presented in the form, $\Psi(t,x) = \psi(x) \exp (-i\mu t)$, where $\mu$ is the chemical potential. In the linear, non-interacting limit, $g_{1D}\to 0$, 
the spectrum of matter waves has the characteristic band-gap structure~\cite{Pearl03}. The nonlinear localization of matter waves in the form of {\em gap solitons} occurs in the gaps of the linear spectrum, and the family of the lowest-order gap solitons in the first (finite) spectral gap is shown in Fig.~\ref{fig_g1} as the dependence of the soliton norm, $P=\int \psi^2 dx$, vs. chemical potential.

\begin{figure}
\includegraphics[width=2.5in]{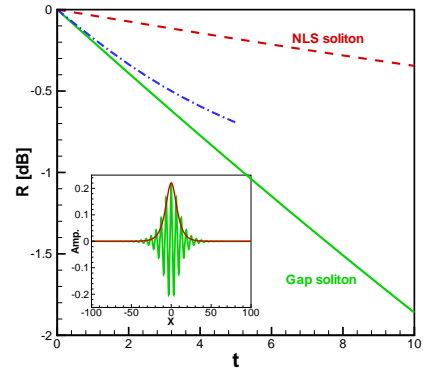}
\caption{(Color online) Time evolution of the optimal squeezing ratio, $R$, for the gap soliton (solid), its near-band-edge envelope (dot-dashed), and the NLS soliton (dashed). Inset: profiles of the gap soliton found from Eq. (\ref{eq_GP}) at $\mu = 1.91$ and its envelope used in the approximation (\ref{NLS}).}
\label{fig-prof}
\end{figure}

Since the degree of localization of gap solitons varies across the gap [see Fig.~\ref{fig_g1}(a-c)], near the bottom edge of the gap, $\mu\approx \mu_0$, the weakly localized soliton profile is well described by the ``envelope'' approximation \cite{pu}, \begin{equation}
\label{NLS} \psi(x; \mu) = A F(x) \Phi(x; \mu_0),
\end{equation}
where $\Phi(x;\mu_0)$ is the periodic Bloch state at the corresponding band edge, and $F(x)$ is a slowly varying function. 
The inset in Fig.~\ref{fig-prof} shows the oscillating wavefunction $\psi(x)$ of a gap soliton near the band edge, at $\mu = 1.91$, together with the corresponding Bloch-wave envelope $F(x)$.
The envelope function $F(x)$ is the solution of the lattice-free GP equation or nonlinear Schr{\"o}dinger (NLS) equation with the effective {\em anomalous} diffraction and interaction energy modified by the lattice \cite{pu}.

To study the quantum fluctuations of the gap solitons, we replace the `classical' mean field described by Eq. (\ref{eq_GP}) by the bosonic field operator $\hat{\Psi}$.
Then, we use the linearization approach with the perturbed quantum field operator $\hat{\psi}$ around the mean-field
solution $\Psi_0$.
By using the {\em back-propagation method}~\cite{Lai95}, we calculate the optimal squeezing ratio for gap solitons as a function of the evolution time.
The optimal squeezing ratio of the quadrature field for gap solitons, as a result of the homodyne detection scheme, can be defined as \cite{Lai95}: $R(t) = \text{min}[{\rm var} \langle \Psi_L(t)|\hat{\psi}(t)\rangle /
{\rm var} \langle \Psi_L(t)|\hat{\psi}(0)\rangle]$, 
where the inner product between the operator field, $\hat{\psi}$, and the local oscillator profile, $\Psi_L$, is defined as:
\[
\langle \Psi_L| \hat{\psi}\rangle=\frac{1}{2}\int \left( \Psi^*_L \hat{\psi}+\Psi_L\hat{\psi}^\dagger \right)dx,
\]
where we use the mean-field soliton with an adjustable phase shift as the local oscillator profile, $\Psi_L = \Psi_0 e^{i\theta}/P$.
By varying the phase of the homodyne detection, $\theta $, we can determine the minimun value of the squeezing ratio of the quadrature component of the quantum field.
When $\theta =0$, the in-phase quadrature component is detected, and when $\theta =\pi /2$, the out-of-phase quadrature
component is detected. 
The result of the calculations of $R(t)$ for the gap soliton near the bottom edge of the spectral gap is presented in Fig.~\ref{fig-prof} by a solid curve.  

The optimal squeezing ratio for gap solitons can be compared with that for a conventional NLS soliton coinciding with the soliton envelope $F(x)$ and evolving in a lattice-free anomalous diffraction regime (dashed line in Fig. \ref{fig-prof}).
The NLS soliton of the same envelope is only weakly quadrature squeezed during propagation due to the nonlinearity of the matter-wave, whereas the gap soliton shows {\em enhanced quantum noise squeezing}.
In contrast, the near-band-edge envelope approximation of Eq. (\ref{NLS}) (dot-dashed line in Fig.~\ref{fig-prof}) provides a good estimate for soliton squeezing in the initial stage of evolution.
\begin{figure}
\includegraphics[width=8cm]{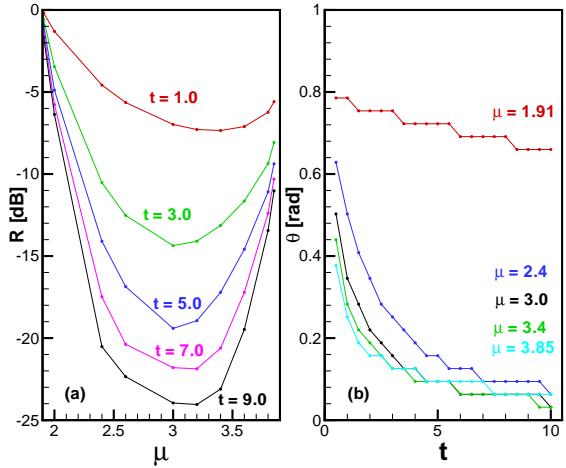}
\caption{(Color online) (a) Optimal squeezing ratio, $R$, for the gap solitons with different chemical potentials, $\mu$, at different time. (b) Time evolution of the phase in the homodyne detection for the optimal squeezing ratios at different values of the chemical potential within the gap. }
\label{fig_g2}
\end{figure}
\begin{figure}
\includegraphics[width=3.5in]{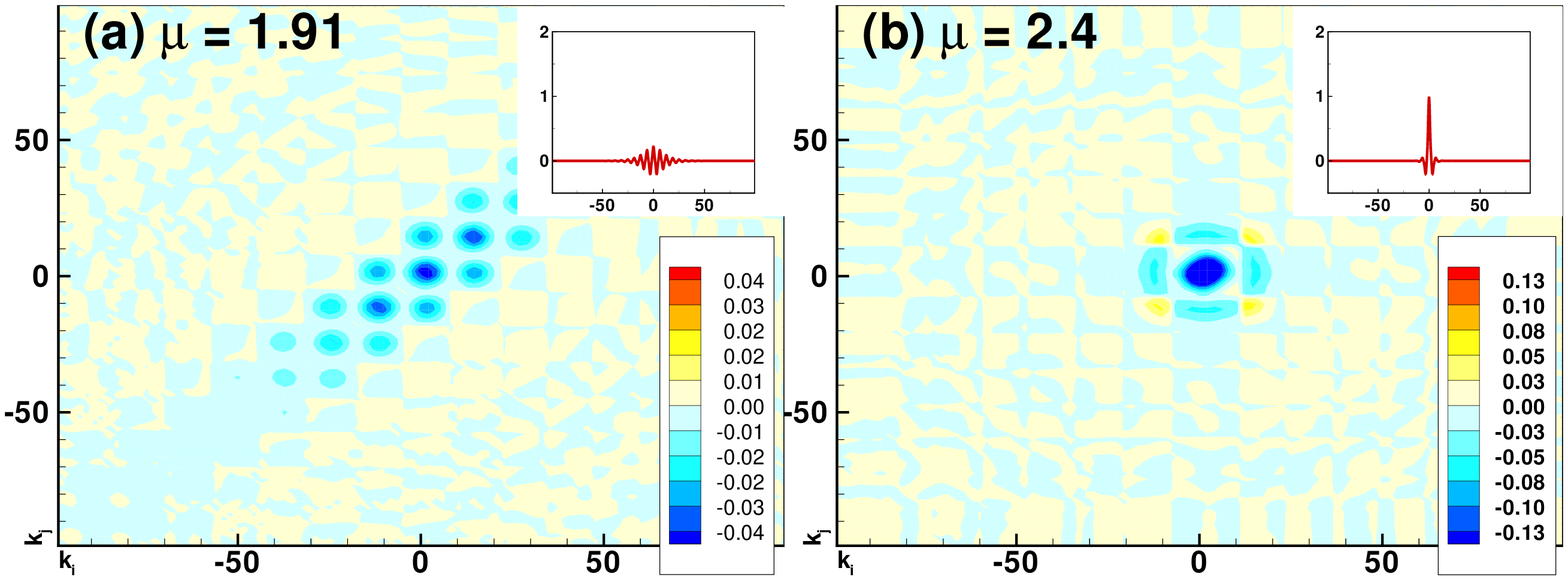}
\includegraphics[width=3.5in]{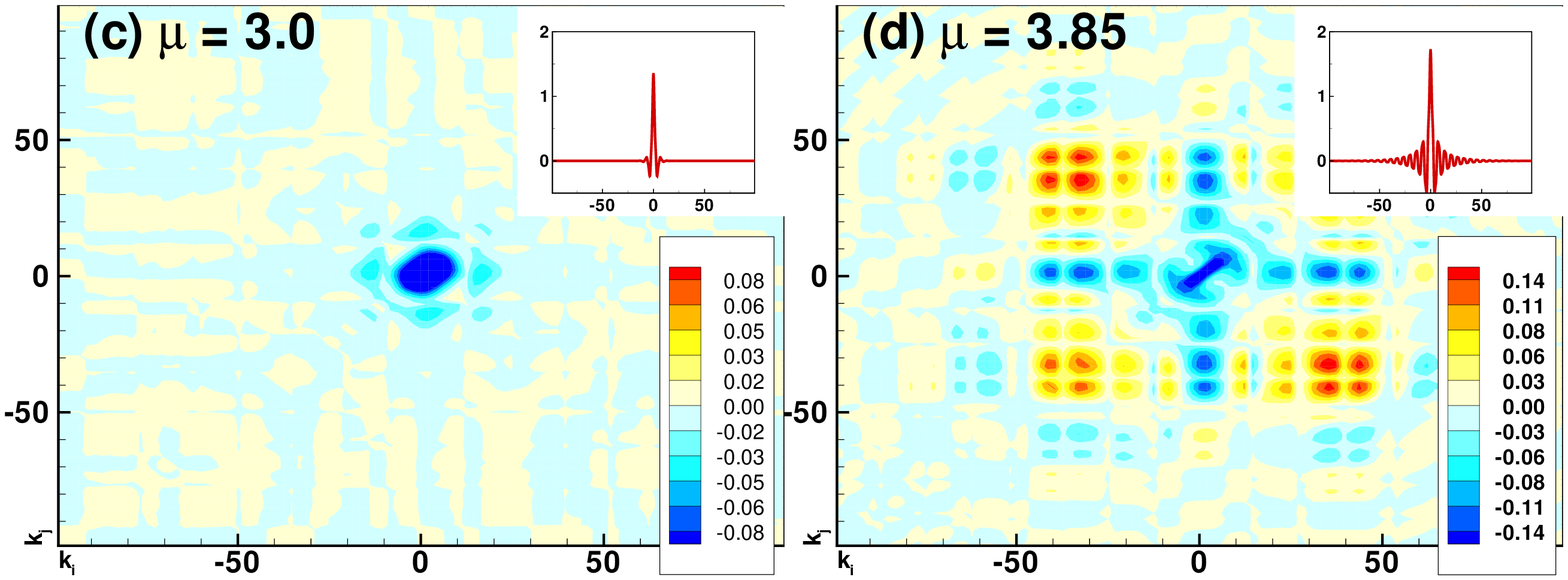}
\caption{(Color online) Quantum correlation spectra in the {\it spatial} ($x$)-domain for gap solitons at different points within the gap and at the values of $\theta$ corresponding to optimal quadrature squeezing: (a) $\theta=41.4^\circ$, (b) $\theta=9^\circ$, (c) $\theta=5.4^\circ$, and (d) $\theta=5.4^\circ$. Insets show the corresponding soliton components in the $x$-domain. }
\label{fig-spck}
\end{figure}

When the chemical potential of the gap soliton moves away from the band edge, the envelope approximation (\ref{NLS}) becomes invalid. In Fig.~\ref{fig_g2}(a), we show the dependence of the optimal squeezing ratio of a gap soliton at different values of the chemical potential inside the gap for different times.
Towards the middle of the gap, the squeezing ratio improves with the chemical potential as the peak density of the gap soliton and hence the nonlinearity are increasing.
Near the band edge the quantum noise of a gap soliton is squeezed in the quadrature field component, Fig.~\ref{fig_g2}(b).
Similar to the amplitude-squeezed Bragg solitons \cite{RK-fbg}, the maximum squeezing angle approaches in-phase quadrature $\theta \to 0$, when the chemical potentials of gap solitons move deeper inside the gap.
However, number-squeezing is never achieved  because we don't use an optimal profile for the local oscillator. For a fixed time and varying chemical potential, the best quadrature squeezing occurs around the middle of the gap.

To understand the band-gap effect on the quantum fluctuation of matter-wave gap solitons, we analyze the quadrature correlations between different spectral components of the gap soliton induced by its nonlinear evolution. The intra-soliton correlation coefficients, $C_{ij}$, are found by calculating the normally-ordered covariance,
\begin{equation}
C _{ij}\equiv \frac{\langle :\Delta \hat{n}_{i}\Delta
\hat{n}_{j}:\rangle}{\sqrt{\Delta \hat{n}_{i}^{2}\Delta
\hat{n}_{j}^{2}}}~,
\label{C}
\end{equation}
where $\Delta \hat{n}_{j}$ is the atom-number fluctuation in the $j$-th slot $\Delta s_{j}$ in the spatial ($s = x$) or momentum ($s = k$) domain:
\[
\Delta \hat{n}_{j}=\int_{\Delta s_{j}}d\,s[\Psi_{L}(t,s)\hat{\psi}^{\dag}(t,s)+\Psi_{L}^{\ast }(t,s)\hat{\psi}(t,s)],
\]
The corresponding correlation spectra in the {\it spatial} ($x$)-domain for gap solitons at different points within the gap and for values of $\theta$ corresponding to the maximum squeezing are shown in Figs.~\ref{fig-spck}(a-d), respectively. The near-band-edge gap soliton has a discrete correlation pattern ``shaped'' by the periodicity of the lattice, with the spatial components being weakly anti-correlated or uncorrelated [Fig. \ref{fig-spck}(a)]. The anti-correlation of the gap soliton components explains its enhanced quadrature squeezing compared to the NLS soliton without a lattice.

In the middle of the gap, the soliton is strongly spatially localized, and its spatial components are strongly anti-correlated [Fig. \ref{fig-spck}(b,c)], which enhances the quadrature noise squeezing.
However, near the top edge of the gap, the localization of the gap soliton degrades due to resonance with the Bloch state at the corresponding edge, and strongly correlated, noisy components associated with the periodic Bloch wave structure start to dominate in the correlation pattern  [Fig. \ref{fig-spck}(d)].
The squeezing is reduced near that edge of the spectral gap.

The possibility of number squeezing for solitons can be assessed by analyzing the intra-soliton number correlations at $\theta=0$. In the {\it momentum} domain NLS soliton displays the well-known symmetric correlation pattern with the noisy, strongly correlated outer regions of the soliton spectrum \cite{levand}.
For this reason, an efficient number squeezing of the NLS solitons can be produced by spectral filtering that removes the noisy spectral components \cite{Friberg}.
On the contrary, the number correlation spectrum of a near-band-edge gap soliton displays a periodic pattern with regions of strong correlations at the Bragg condition,  $k_i=k_j = \pm (2m+1)$ ($m$ is an integer) and uncorrelated off-diagonal regions. This correlation pattern persists through the entire gap until the noisy spectral components due to resonance with a Bloch state appear near the opposite band edge.
Therefore the enhancement of number squeezing of the BEC gap soliton {\em with} additional spectral filtering seems unfeasible due to spectral selectivity of the Bragg scattering.

The possibility to achieve squeezing of atomic solitons paves the way to creating entangled macroscopic coherent states in BEC interference experiments. However, optical lattices also offer an alternative possibility for entanglement production.
It is known that the optical lattice supports stationary states in the form of bound soliton pairs~\cite{Pearl03}.
The nonlinear interaction between solitons in a pair could induce entanglement and hence nonseparability of the bound state. This entanglement can potentially be exploited in BEC atom number detection by implementing analogs of quantum nondemolition measurements based on entangled pairs of optical solitons \cite{non-dem}. Potential advantage of gap solitons compared to the lattice-free atomic solitons is the long-lived nature of the bound states which are pinned by the lattice and do not break-up as a result of evolution.  Hence strong interaction-induced atom-number correlations can develop between the atomic wavepackets.
 \begin{figure}
\includegraphics[width=3.0in]{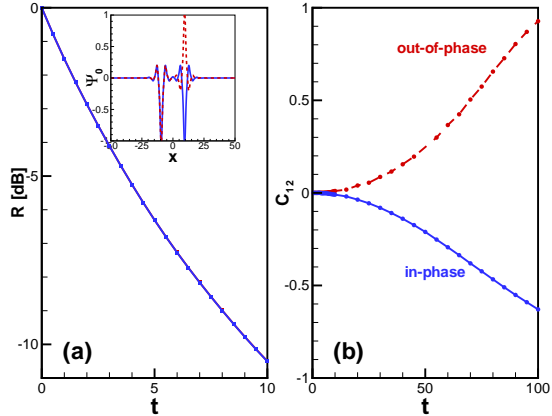}
\caption{(Color online) (a) Optimal squeezing ratio for in-phase (inset, solid) and out-of-phase (inset, dashed) bound soliton states ($V_0 = 4.0$, $\mu = 2.5$); the two curves overlap. (b) Atom number correlation $C_{12}$ for in-phase (solid) and out-of-phase (dashed) soliton pairs.}
\label{fig_b1}
\end{figure}

Typical in-phase and out-of-phase bound states of two gap solitons in the middle of spectral gap are shown in Fig.~\ref{fig_b1}. These states contain the same number of atoms for the same $\mu$, which equals to exactly twice the number of atoms in a single soliton.
Remarkably, the {\em internal} noise correlation properties are the same for both pairs, which results in their identical optimal number squeezing ratio,  see Fig.~\ref{fig_b1}(a).
Soliton entanglement can be quantified by inter-soliton atom number correlations which can be calculated according to Eq. (\ref{C}), with $\Delta \hat{n}_{i,j}$ being perturbations of the atom number operator of two solitons, and indices $\{i,j\}=\{1,2\}$ numbering individual solitons.  The evolution of the correlation parameter $C_{12}$ shows dramatic difference between the in- and out-of-phase soliton pairs. As follows from Fig.~\ref{fig_b1}(b), the atom number correlation parameter for the in-phase soliton pair is negative while it is positive for the out-of-phase pair. The difference in the entanglement behavior can be understood by examining the interaction energy of the solitons in the pair $E_{12}=\int {\cal{E}}_{12}(x) dx$, with the energy density:
\[
{\cal{E}}_{12}=\sum_{i\neq j}(\frac{1}{2} \frac {d \psi_{i} }{dx}\frac {d \psi^*_{j} }{dx}+V(x)\psi_{i}\psi_{j}+3|\psi|^2_{i}|\psi|^2_{j}+4\psi^3_{i}\psi_{j}).
\]
Interaction energy, $E_{12}$ depends on overlapping tail structure of the stationary profiles $\psi_{1,2}(x)$ corresponding to mean-field solitons $\Psi_0$. This quantity is small but positive, $E^{\rm out}_{12}=3.1\times 10^{-2}$, for the out-of-phase soliton pair, and negative, $E^{\rm in}_{12}= - 3.4\times 10^{-2}$, for the in-phase pair. This results in positive and negative atom-number correlation parameter for the respective pairs. Small difference in the interaction energies is responsible for the slight asymmetry in correlations development seen in Fig.~\ref{fig_b1} (b). As a result, longer evolution times are required for the development of maximum atom number correlations between the solitons in an in-phase pair.

In conclusion, we have investigated the effect of the periodic potential on quantum fluctuations and entanglement properties of  gap solitons in repulsive BEC confined by an optical lattices.
We have revealed that the quadrature squeezing of gap solitons can be significantly enhanced compared to the matter-wave solitons in the lattice-free case. This enhancement is a strong effect that can potentially be detected in experiments on BEC gap solitons in one-dimensional optical lattices.
We would like to emphasize that the basic results of our analysis can be useful in the study of the bandgap effects on the quantum squeezed states in other fields, such as quantum optics and gap solitons in photonic crystals.

The work was supported by the Australian Research Council.
We thank P. Drummond, H. Bachor, G. Kurizki and P.-K. Lam for useful discussions.

\end{document}